\documentclass[conference]{IEEEtran}
\IEEEoverridecommandlockouts
\usepackage{cite}
\usepackage{amsmath,amssymb,amsfonts}
\usepackage{algorithmic}
\usepackage{graphicx}
\usepackage{textcomp}
\usepackage{xcolor}
\def\BibTeX{{\rm B\kern-.05em{\sc i\kern-.025em b}\kern-.08em
    T\kern-.1667em\lower.7ex\hbox{E}\kern-.125emX}}
\begin{document}

\title{Automated SVA Generation with LLMs

\author{
\IEEEauthorblockN{
Lik Tung Fu\textsuperscript{1,2},
Qihang Wang\textsuperscript{2},
Shaokai Ren\textsuperscript{2},
Mengli Zhang\textsuperscript{2},
Sichao Yang\textsuperscript{3},
Jun Liu\textsuperscript{3},
Xi Wang*\textsuperscript{1,2}
}
\IEEEauthorblockA{\textsuperscript{1}National ASIC Center, School of Integrated Circuits, Southeast University, China}
\IEEEauthorblockA{\textsuperscript{2}National Center of Technology Innovation for Electronic Design Automation, China}
\IEEEauthorblockA{\textsuperscript{3}XEPIC Corporation Limited, China}
\IEEEauthorblockA{
Email: liktungfu@seu.edu.cn, worlesenric@gmail.com, renshaokai@nctieda.com, zhangmengli@nctieda.com,\\
sichaoy@x-epic.com, brightl@x-epic.com, xi.wang@seu.edu.cn
}
\IEEEauthorblockA{*Corresponding author.}
}
\thanks{This work is supported by the Key Research and Development Program of Jiangsu Province (Grant No.\ BG2024010), the National Natural Science Foundation of China (NSFC Grant No.\ 92464301), and the National Key Research and Development Program (Grant No.\ 2024YFB4405600).}

}

\maketitle

\begin{abstract}
Functional verification remains a dominant cost in modern IC development, and SystemVerilog Assertions (SVAs) are critical for simulation-based monitoring and formal property checking. However, writing SVAs by hand is time-consuming and error-prone. Directly prompting general-purpose large language models (LLMs) is also unreliable: the generated properties are often syntactically invalid or semantically incorrect, and the problem is exacerbated by scarce, high-quality domain training data. We present SVA Generator, a data-centric framework that translates natural-language SVA Descriptions (SVADs) into executable SVAs. It uses AST-grounded constraint injection and an automated supervision pipeline that enforces structural consistency and reduces hallucinations via de-duplication and constraint checks. To enable rigorous evaluation, we introduce a benchmark suite stratified by AST depth and use formal property equivalence checking to quantify semantic correctness separately from syntax validity, by checking mutual implication between the generated and reference properties under the same clocking and environment assumptions. Across all difficulty tiers, SVA Generator achieves comparable Syntax Pass Rate (SPR) to strong general LLM baselines, while delivering substantially higher Semantic Equivalence Rate (SER) on deeper tiers: +24.5 pp on D2, +26.0 pp on D3, and +17.5 pp on D4 relative to the best-performing general LLM, corresponding to a +22.7 pp SER improvement on average over D2--D4. These results highlight that high-fidelity data construction and depth-stratified benchmarking are key to reliable, semantics-preserving SVA generation.

\end{abstract}


\section{Introduction}
\label {sec:introduction}
The slowing benefits of Moore's law have shifted progress away from straightforward transistor scaling. Instead, designers increasingly rely on architectural innovation, specialization, and heterogeneous integration. These trends continue to drive performance gains, but they also increase system-level complexity. For example, designs integrate more IP blocks and richer interconnect fabrics. They also introduce multiple clock and power domains, as well as tighter hardware/software coupling. Therefore, the functional state space expands substantially, and more behaviors must be validated. As a result, functional verification often consumes more than half of the overall development effort and becomes a dominant bottleneck in time-to-market~\cite{simens_report_2023}.
In this context, Assertion-Based Verification (ABV) has been widely adopted~\cite{surveyassertion}. SVAs can serve as executable specifications, enabling simulation-based dynamic verification. They also support exhaustive property checking in formal verification. Consequently, SVAs improve observability, promote maintainable verification collateral, and enable earlier bug discovery compared with purely testbench-driven approaches.

Nevertheless, the practical adoption of ABV is constrained by the cost of authoring high-quality SVAs. SVAs combine concurrent semantics with linear temporal logic, and their correct usage requires expert knowledge of clocking, reset conventions, and subtle temporal operators. Moreover, each SVA typically captures only a single property, which makes large-scale manual development labor-intensive and error-prone. Therefore, automating SVA generation is highly desirable to reduce the expertise barrier and to scale ABV to industrial designs~\cite{hasan2015formal,revolution}.

Recent progress in large language models (LLMs) has demonstrated strong performance in general code generation~\cite{chatcpu,iDSE,rechisel,location,vgv}, but directly applying general-purpose LLMs to SVA generation remains unreliable. SVA generation requires long-chain reasoning over cycle-accurate temporal dependencies under strict syntactic constraints, rather than shallow translation. Under limited domain supervision, general LLMs often produce SVAs that compile but deviate semantically from the intended property, resulting in ``syntax-correct yet functionally wrong'' SVAs. This issue is exacerbated by the scarcity of high-quality, domain-specific SVAD--SVA pairs that precisely align natural-language intent with formally checkable code. Finally, evaluation is frequently restricted to surface-level metrics and lacks standardized protocols for semantic correctness at scale, which hinders reproducible comparison and systematic progress~\cite{FIXME,genben1,assertllm,chatsva}.

To address these issues, we propose a framework for reliable SVA generation, treating it as a foundational subtask for automated SVA development. Our approach is data-centric: instead of relying on unconstrained text generation, we leverage Abstract Syntax Tree (AST)-grounded structural constraints and formal verification feedback to construct reliable supervision and to improve generation robustness. By aligning data construction, training, and evaluation within a unified pipeline, the proposed framework targets both syntactic validity and semantic equivalence.

Our main contributions are summarized as follows:
\begin{itemize}
    \item We propose an SFT-based framework for generating executable SVAs from natural-language SVA Descriptions.
    \item We present a data-centric pipeline that automatically constructs high-quality supervision for SVA generation, enabling scalable training with reduced hallucinations.
    \item We introduce a depth-stratified benchmark suite with formal property equivalence checking for rigorous evaluation, which will be released publicly upon publication.
\end{itemize}


\section{Motivation}
\label{sec:background}

\subsection{Background}
SVAs are a cornerstone of ABV because they encode temporal intent as executable properties that can be monitored in simulation and proven by formal verification. An SVA is typically built from sequences and properties: sequences specify ordered event patterns across clock cycles, while properties constrain how sequences trigger, imply, or coincide with other behaviors. To express realistic verification intent, the language provides a rich set of temporal operators, including implication, cycle delays, repetition ranges, sampling functions, and local variables for binding intermediate values. This expressiveness comes with semantic subtleties, including the distinction between overlapping and non-overlapping implication, vacuity, reset semantics, and deep nesting of sequential composition. As a result, correct SVA authoring demands cycle-accurate reasoning under strict syntactic and semantic constraints, and the space of valid implementations is significantly narrower than that of conventional software code generation.

Automating SVA generation has therefore been studied for years. Early \emph{dynamic} approaches mined candidate SVAs from simulation traces~\cite{harm,ateam,goldmine}. Although trace-driven mining can produce useful monitors, it fundamentally depends on the testbench and the observed behaviors, and thus cannot guarantee coverage of unexercised corner cases. Moreover, when the design-under-test contains functional defects, mined properties may inadvertently encode buggy behavior, weakening their value for verification. In parallel, \emph{static} approaches used traditional NLP to parse user-provided descriptions into structured templates and then applied rule-based mappings to SVA syntax~\cite{GLAsT-NLP,subtreeNLP,CNLNLP,AutomatedNLP}. Later studies incorporated hybrid learning components for entity extraction, template selection, and signal mapping~\cite{ChatbotNLP,HybridNLP,SpectosvaNLP} to relax rigid input constraints. However, pre-LLM NLP pipelines were limited by shallow semantic understanding and brittle template assumptions, and they often failed under paraphrasing, long-range dependencies, or nontrivial timing relations.

Recent advances in LLM-based code generation have motivated a new wave of work on LLM-assisted EDA and hardware design/verification tasks~\cite{meic,location,uvllm,chatchisel}. Existing LLM-aided SVA generation efforts can be broadly divided into two application directions. The first direction continues the classical NLP line of research: generating SVAs from precise, localized natural-language descriptions, aiming to improve temporal reasoning fidelity and code realization~\cite{Usingllm,securityllm,validatablellm,finetunellm,towards,domainllm}. The second direction targets higher-level automation: extracting SVAs from more abstract artifacts such as specifications, verification plans, or other design documents, often by decomposing the workflow into multiple LLM-driven steps or agents~\cite{chatsva,assertllm,AssertionForge,chiraag}. While this document-level direction is practically appealing, it also couples multiple hard subproblems, including requirement understanding, signal resolution, property selection, and property implementation, which makes errors harder to diagnose and evaluation less controlled~\cite{chipmind}. In contrast, this paper focuses on a classical yet fundamental task: translating a precise SVA Description (SVAD) into an executable SVA. This formulation isolates the core challenge of temporal reasoning and code realization, enables controlled evaluation, and provides a reusable component that can be integrated into broader verification pipelines without requiring full document-level understanding.

\begin{figure*}[t]
    \centering
    \includegraphics[width=\textwidth]{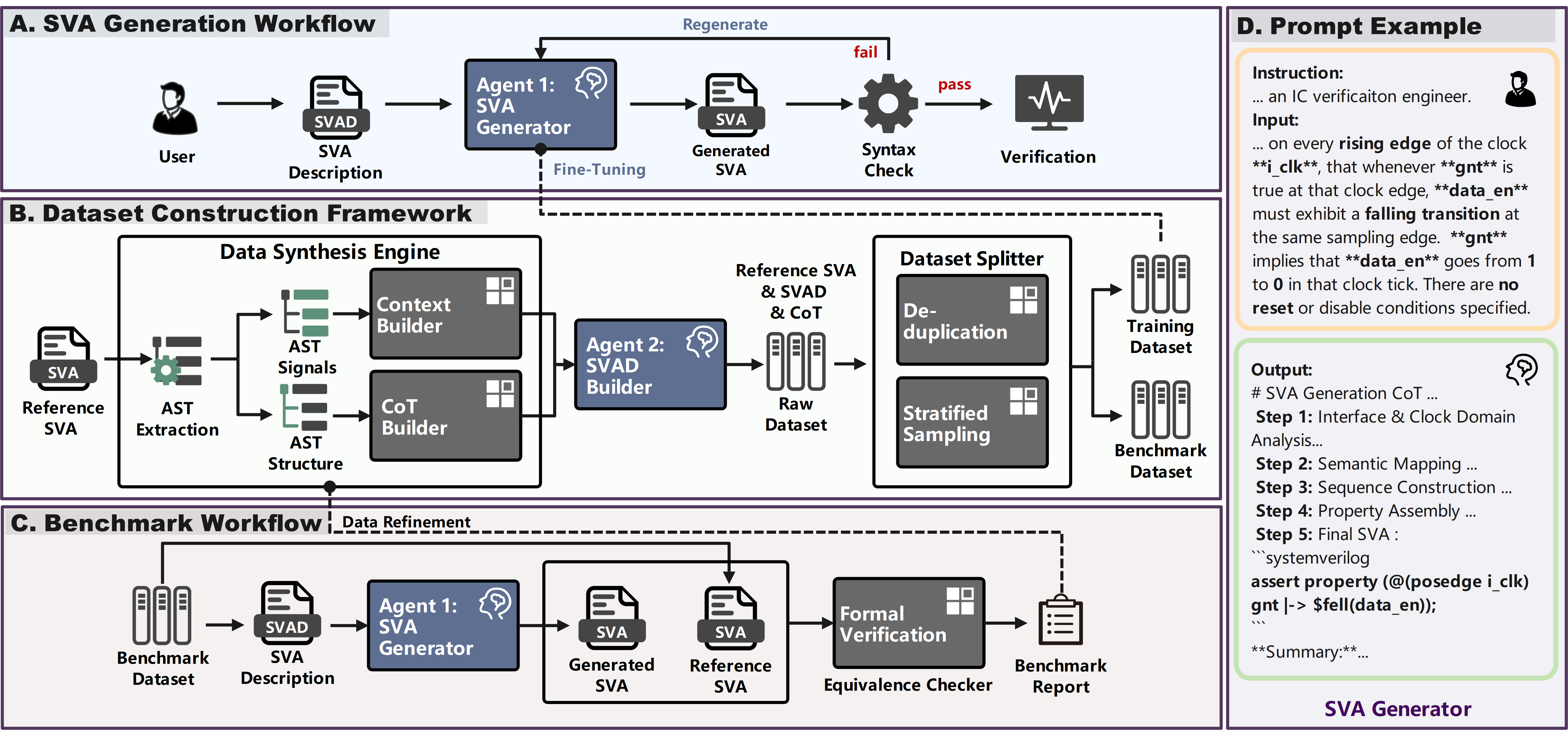}
    \caption{Overall architecture. (A) Iterative SVA generation workflow with syntax-aware regeneration. (B) AST-grounded dataset construction framework producing aligned SVA, SVAD and CoT. (C) Benchmark workflow with formal property equivalence checking and data refinement feedback. (D) Prompt example illustrating the SVAD input format and the generated CoT and SVA output.}
    \label{fig:overall}
\end{figure*}

\subsection{Challenges}
Despite the narrowed scope, two key challenges remain and motivate our design.

\textbf{C1: Data scarcity.}
High-quality training pairs that align SVADs with correct SVAs are scarce. Industrial SVAs are typically proprietary, and public repositories are limited in both scale and diversity. More critically, even when SVA code is available, it rarely includes accurate and unambiguous natural-language descriptions. Naively generating descriptions with an LLM can introduce hallucinated signals or distorted timing relations, which can corrupt supervision and directly degrades downstream SFT. Therefore, scalable dataset construction must be grounded in the code structure, and it must enforce strict consistency constraints to prevent semantic drift.

\textbf{C2: Insufficient semantic evaluation.}
Many existing benchmarks emphasize surface-level criteria such as compilability or rely on heuristic similarity measures, which are inadequate for temporal logic. For SVAs, syntactic validity is necessary but not sufficient: two SVAs can compile while specifying different behaviors due to a single operator choice or an incorrect timing window. Consequently, evaluation must directly measure semantic correctness via formal methods whenever possible, and it should stratify difficulty to expose how performance degrades with deeper temporal nesting and longer reasoning chains.

\section{Methodology}
\label{sec:Design}

To address \textbf{C1}, we construct a high-quality dataset using AST-grounded constraint injection and an automated supervision pipeline that enforces structural consistency. To address \textbf{C2}, we introduce a depth-stratified benchmark and evaluate semantic correctness via formal property equivalence checking based on mutual implication under consistent clocking and environment assumptions. Fig.~\ref{fig:overall} summarizes the resulting end-to-end pipeline.

\subsection{The SVA Generation Workflow}
As illustrated in Fig.~\ref{fig:overall}(A), the workflow takes a user-provided SVAD as input. An SVAD is a concise natural-language description of the intended temporal property, including its trigger, expected response, and timing constraint. By focusing on SVA generation, the workflow isolates the core task of converting temporal intent into executable SVA.

Our inference engine is \texttt{Agent1} (SVA Generator), an LLM specialized for SVA synthesis through SFT on the high-quality dataset constructed in Section~\ref{sec:dataset}. Given an SVAD, \texttt{Agent1} generates a candidate SVA in standard SystemVerilog syntax while following ABV coding conventions, including proper property/sequence structure, explicit clocking, and well-formed temporal operators. Fig.~\ref{fig:overall}(D) presents a representative prompt and its outputs, including both an intermediate CoT and the final SVA.

To improve reliability, we incorporate a syntax-aware refinement loop. Each generated SVA is first passed to a Syntax Checker for parsing and compilation validation. If the check succeeds, the SVA is returned as the final output for downstream simulation or formal verification. Otherwise, the compiler error log is fed back to \texttt{Agent1} and the SVA is regenerated. This process repeats until the output satisfies the syntax constraints, reducing invalid outputs without manual debugging.

The resulting output is a syntactically valid SVA that preserves the intent expressed in the SVAD and is ready for integration into ABV environments.

\subsection{Dataset Construction Framework}
\label{sec:dataset}

Fig.~\ref{fig:overall}(B) illustrates the dataset construction framework. To address \textbf{C1}, it builds an automated high-quality synthesis pipeline that produces reliable supervision for SFT while reducing inconsistent annotations. Starting from a corpus of \emph{reference SVAs}, the framework constructs aligned $(\text{SVA}, \text{SVAD}, \text{CoT})$ entries, which are then partitioned into training and benchmark sets.


\subsubsection{Data Synthesis Engine}
The Data Synthesis Engine converts raw SVA code into structured signals for annotation. It begins with static analysis and extraction, where each SVA is parsed into an AST. From the AST, we extract two complementary views. First, AST Signals capture the symbol-level interface required for faithful description, including clocking events, reset/disable conditions when present, referenced design signals, and system functions used by the SVA. Second, AST Structure captures the topology of temporal logic, including the hierarchy of sequence/property composition, operator nesting, and the placement of delays and repetitions. This separation is critical: signals constrain \emph{what} entities may appear in SVADs, while structure constrains \emph{how} these entities are related temporally.

Based on the extracted information, the engine provides two builders that jointly control annotation quality. The Context Builder performs constraint injection by converting AST Signals into a structured prompt context. The injected constraints explicitly enumerate the allowed signal names and key syntactic elements that must be referenced. This forces the annotator to ground SVAD generation in the original SVA, preventing common hallucination modes such as introducing non-existent signals, renaming identifiers, or omitting essential clock/reset semantics. In parallel, the CoT Builder constructs a reasoning trace from AST Structure using deterministic, rule-based derivations. The produced CoT explains the mapping from temporal intent to SVA operators, for example translating ``after $N$ cycles'' into $\#\#N$, or translating implication semantics into $|->$ versus $|=>$. By generating CoT algorithmically rather than sampling it from an LLM, we obtain stable intermediate supervision that is consistent across paraphrases and robust to annotation variance.

\subsubsection{Annotation Agent}
With the injected constraints and deterministic CoT, \texttt{Agent2} (\emph{SVAD Builder}) generates a natural-language SVAD aligned with the input SVA. The Context Builder constrains the allowable signal names and related syntax, while the CoT Builder constrains the temporal structure of the property. Under these constraints, \texttt{Agent2} produces readable and unambiguous descriptions that preserve the intent of the original SVA. The resulting SVA--SVAD--CoT alignments are collected into a raw dataset.


\subsubsection{Dataset Splitter}
To ensure reproducibility and prevent evaluation leakage, the framework includes a Dataset Splitter with two stages. First, we perform \textbf{de-duplication} to remove near-duplicate samples using a strict matching protocol over normalized SVAs and their associated metadata, eliminating repeated patterns that would otherwise inflate reported performance. Second, we perform \textbf{stratified sampling} to partition the cleaned dataset into a training dataset and a benchmark dataset. Stratification is performed jointly by AST depth $D$ and SVA syntax categories, so that the benchmark set remains distributionally comparable to training while preserving a controlled coverage of structural complexity. This split supports fair evaluation of \texttt{Agent1} on unseen dataset and enables difficulty-aware analysis in Section~\ref{sec:benchmark}.

\subsection{Benchmark Suite \& Evaluation Protocol}
\label{sec:benchmark}
Fig.~\ref{fig:overall}(C) illustrates the benchmark workflow and evaluation protocol. To address \textbf{C2}, we establish a reproducible benchmark suite with semantic correctness checking and difficulty-aware reporting.


\subsubsection{Benchmark composition}
The Benchmark Dataset is produced as a hold-out branch from the Dataset Splitter in Fig.~\ref{fig:overall}(B). It is drawn from the same source distribution as the training data, yet is strictly isolated to ensure fair generalization evaluation. To quantify structural difficulty, we adopt \textbf{AST depth} $D$ as the primary complexity metric, defined as the maximum depth of the AST generated from an SVA. A larger $D$ indicates deeper nesting of temporal operators and longer reasoning chains, which is known to challenge both LLMs and traditional parsers. Based on $D$, we stratify the benchmark into four difficulty tiers, D1--D4, which disentangles SVA syntax coverage from structural complexity and enables fine-grained analysis across increasing temporal nesting.

Beyond stratification, the suite is designed to be syntactically comprehensive. The benchmark covers a broad spectrum of SVA constructs, including implication, cycle delays, repetition operators, and local variables, thereby ensuring that performance trends are not dominated by a narrow subset of patterns. This completeness is crucial for ABV usage, where practical SVAs often combine multiple operators within a single property.

\subsubsection{Evaluation workflow}
Given a benchmark SVAD, we perform inference with \texttt{Agent1} using frozen parameters to generate a candidate SVA, as shown in Fig.~\ref{fig:overall}(C). Each candidate is then paired with its reference SVA and passed to a formal property equivalence checker. The checker evaluates semantic correctness by checking mutual implication between the generated and reference properties under the same clocking and environment assumptions, and it further categorizes non-equivalent cases into directional relations such as being weaker or stronger than the reference. This semantic evaluation complements syntax-based validation and directly measures the property-level correctness that ABV requires.

\subsubsection{Benchmark report and data refinement}
The evaluation results are aggregated into a Benchmark Report that summarizes performance by difficulty tier and by failure mode. Importantly, as indicated by the dashed feedback path in Fig.~\ref{fig:overall}(C), the report also serves as a data refinement signal for the dataset construction framework in Section~\ref{sec:dataset}. In particular, systematic failures on higher tiers such as D4 expose missing patterns or insufficient supervision in the synthesized data; these cases are fed back to adjust the Data Synthesis Engine, for example by increasing the sampling weight of underrepresented structures or strengthening constraint injection rules. This closes an iterative improvement loop that progressively aligns the training distribution with the semantic difficulties revealed by formal evaluation.

\section{Implementations}
\label{sec:impl}
\begin{figure*}[t]
    \centering
    \includegraphics[width=\textwidth]{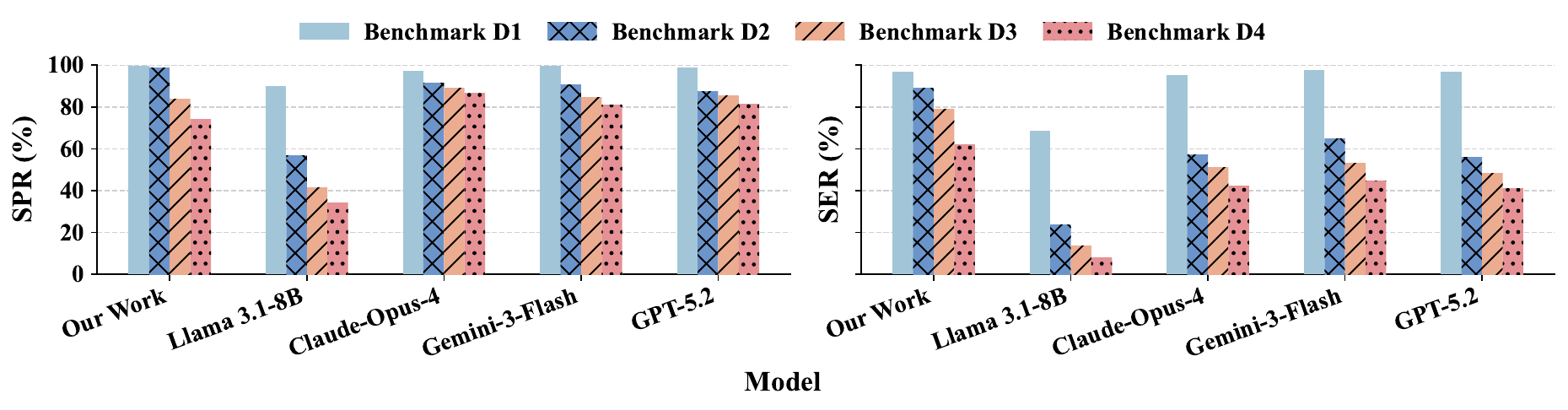}
    \caption{Single-pass comparison across difficulty tiers (D1--D4). Left: SPR. Right: SER. Results are averaged over five runs.}
    \label{fig:main}
\end{figure*}

\subsection{Experimental Setup and Toolchain}
We implement \texttt{Agent1} with Llama-3.1-8B as the base model, balancing code-generation capability and deployment efficiency. For data annotation, \texttt{Agent2} uses GPT-5.2 to generate high-quality SVADs under AST-grounded constraints, while the CoT Builder deterministically derives CoT from the AST structure.
All training runs are conducted on an 8$\times$A800-SXM4-80GB cluster with DeepSpeed for distributed training.
We adopt a two-stage training protocol, continued pre-training followed by SFT: the model is first adapted on domain text including IP specifications and SVA syntax materials, and then fine-tuned on the curated SVAD--SVA pairs produced by the framework in Section~\ref{sec:dataset}. Unless otherwise stated, we use a sequence length of 4096, a global batch size of 32, a learning rate of 5$\times 10^{-5}$, and train for 2 epochs.

Our raw SVA corpus is collected from a proprietary industrial repository spanning diverse ABV scenarios, ranging from interface protocol checking to internal finite-state machine properties. We integrate \textbf{Verible}~\cite{verible} as the front-end static analysis tool for AST parsing and normalization. For semantic consistency filtering during data cleaning and for benchmark evaluation, we use the \textbf{X-Epic GalaxFV SVA Evaluator} as the equivalence checking backend. The final dataset contains approximately 8$\times 10^{4}$ samples and is split into training and benchmark sets with a ratio of 9:1.

\section{Evaluations}
\label{sec:eval}

\subsection{Metrics and Benchmark Stratification}
\label{sec:metrics}

We evaluate SVA generation from two complementary perspectives: syntactic validity and semantic correctness. Results are reported both overall and stratified by benchmark difficulty tiers defined by AST depth.

\noindent\textbf{Syntax Pass Rate (SPR).}
SPR measures the fraction of generated SVAs that can be successfully compiled by a SystemVerilog toolchain. We define $\mathrm{SPR}=N_{\text{pass}}/N_{\text{total}}$, where $N_{\text{total}}$ is the number of benchmark samples and $N_{\text{pass}}$ is the number of outputs that pass the syntax checker. SPR affects the efficiency of the syntax-aware regeneration loop in Fig.~\ref{fig:overall}(A): higher SPR generally means fewer regeneration rounds, lower latency, and lower token cost. Although our framework includes a syntax-aware regeneration loop, we report single-pass SPR and SER in evaluation to ensure a fair comparison with general-purpose LLM baselines that do not use iterative correction.

\noindent\textbf{Semantic Equivalence Rate (SER).}
SER measures semantic correctness using formal property equivalence checking between the generated SVA and the reference SVA. Because syntactic validity does not imply functional correctness, SER is the primary metric for formal semantic evaluation. We compute SER over syntax-valid outputs as $\mathrm{SER}=N_{\text{eq}}/N_{\text{pass}}$, where $N_{\text{eq}}$ is the number of syntax-valid generated SVAs proven equivalent to their references by the equivalence checker. In practice, a semantically incorrect SVA can mislead debugging in the same way as a design bug; SER therefore captures the most important quality dimension beyond compilation success.

\noindent\textbf{Benchmark stratification by AST depth.}
To expose how performance scales with temporal-logic complexity, we stratify the benchmark by AST depth $D$, defined as the maximum depth of the AST parsed from the reference SVA. We group samples into four tiers: \textbf{D1} ($D{=}1$) for shallow properties with minimal nesting, \textbf{D2} ($D{=}2$) for single-level temporal composition such as fixed-cycle delays, \textbf{D3} ($D{=}3$) for multi-level nesting, and \textbf{D4} ($D{=}4$) for deeply nested properties that typically require long-chain temporal reasoning. This stratification enables difficulty-aware reporting of SPR and SER and supports the error attribution analysis in Section~\ref{sec:error-attribution}.

\begin{figure*}[t]
    \centering
    \includegraphics[width=\textwidth]{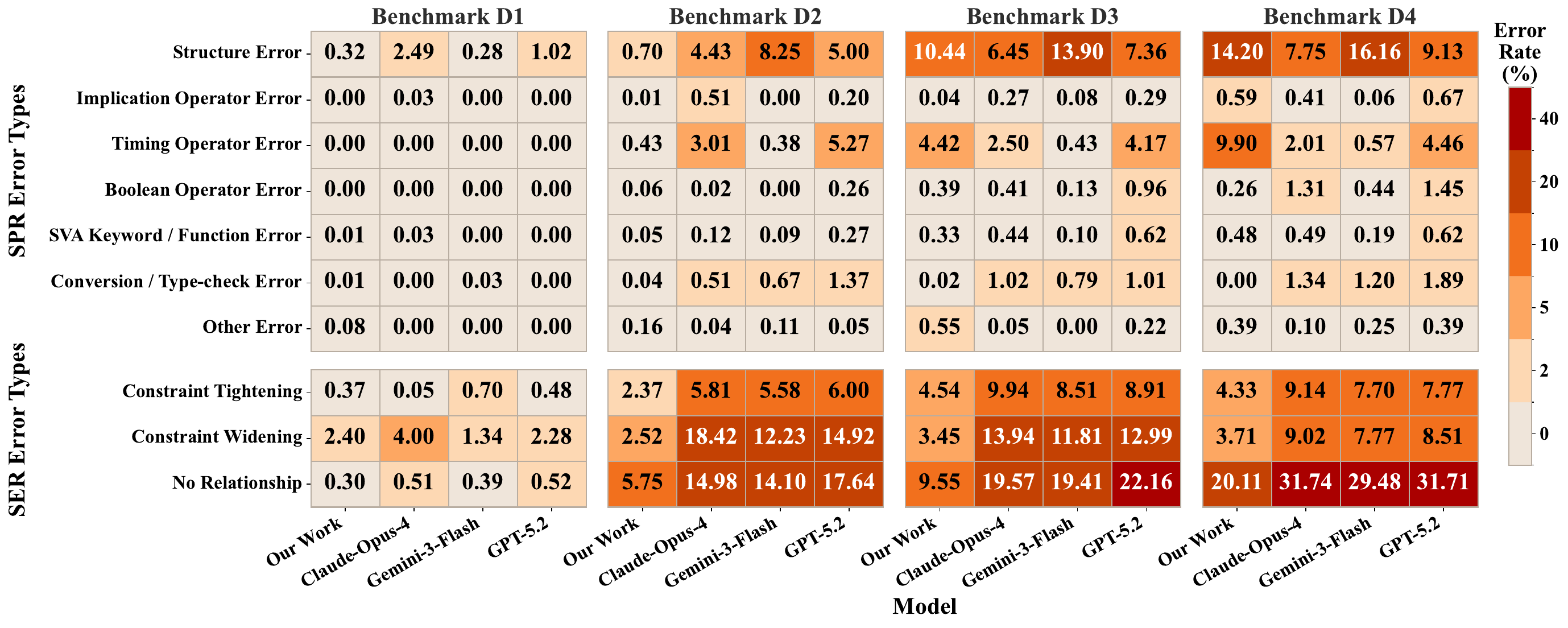}
    \caption{Error attribution heatmaps. Upper: SPR error types over all samples. Lower: SER error types over all samples.}
    \label{fig:err-attr}
\end{figure*}

\begin{table}[t]
\caption{Error labels used in error attribution.}
\label{tab:err-attr-labels}
\begin{center}
\small
\renewcommand{\arraystretch}{1.15}
\begin{tabular}{|p{0.33\linewidth}|p{0.55\linewidth}|}
\hline
\textbf{Label} & \textbf{Description} \\
\hline
Structure Error &
The parser reports missing or unexpected structural elements. \\
\hline
Implication Operator \newline Error &
The error message explicitly mentions implication operators (\texttt{|->}, \texttt{|=>}). \\
\hline
Timing Operator Error &
The error involves delay or repetition constructs (\texttt{\#\#}, \texttt{\#\#[ ]}, \texttt{[->}, \texttt{[=]}, \texttt{[*]}, and related bracket usage). \\
\hline
Boolean Operator \newline Error &
The error is triggered by boolean or relational operators (\texttt{\&\&}, \texttt{||}, and related operators). \\
\hline
SVA Keyword / \newline Function Error &
The error involves an SVA-specific keyword or sampling function, such as \texttt{\$rose}, \texttt{\$fell}, \texttt{\$pass}, \texttt{\$stable}, \texttt{until}, or \texttt{within}. \\
\hline
Conversion / \newline Type-check Error &
The tool reports a conversion, type-check, or mismatch issue, such as: width mismatch; size mismatch; or invalid casts. \\
\hline
Other Error &
The error does not match any of the categories above. \\
\hline
Constraint Tightening &
The generated SVA implies the reference SVA. \\
\hline
Constraint Widening &
The reference SVA implies the generated SVA. \\
\hline
No Relationship &
Neither SVA implies the other. \\
\hline
\end{tabular}
\end{center}
\end{table}

\subsection{Comparative Analysis}
\label{sec:comparative}

Fig.~\ref{fig:main} compares the proposed approach (``Our Work'' in figures) with GPT-5.2, Gemini-3-Flash, Claude-Opus-4, and the base Llama-3.1-8B without task-specific SFT. Each model performs single-pass generation on the benchmark without applying the regenerate loop, and we report the mean SPR and SER over five independent runs. The upper panel shows SPR and the lower panel shows SER, both stratified by AST depth tiers D1--D4.

On SPR, the proposed approach achieves 99.6\%, 98.5\%, 83.8\%, and 74.2\% from D1 to D4, and is broadly comparable to strong general-purpose LLMs. All models show a monotonic SPR drop as depth increases due to the increased syntactic burden of deeply nested SVA constructs. General LLMs are similar on D1--D2 and remain higher on D3--D4.

On SER, the proposed approach achieves 96.9\%, 89.2\%, 79.1\%, and 62.1\% from D1 to D4, attaining the highest SER at every tier. The base Llama-3.1-8B degrades sharply with depth, reaching SER of 23.5\% (D2), 13.7\% (D3), and 7.8\% (D4), highlighting the importance of task-specific SFT for long-chain temporal reasoning and operator selection. Compared with the best-performing general LLM (Gemini-3-Flash), the proposed approach improves SER by \textbf{+24.5} pp on D2, \textbf{+26.0} pp on D3, and \textbf{+17.5} pp on D4, corresponding to a \textbf{+22.7} pp SER improvement on average over D2--D4.

\subsection{Error Attribution}
\label{sec:error-attribution}

Fig.~\ref{fig:err-attr} summarizes error attribution using two heatmaps. The upper heatmap attributes SPR failures by syntax-error types, and the lower heatmap attributes SER outcomes by implication relationships. The definitions of all error types and relationship labels used in this section are summarized in Table~\ref{tab:err-attr-labels}.

\subsubsection{SPR error attribution}
Structure Error is the dominant contributor for all models across D1--D4, and its rate increases with AST depth, which matches the growing syntactic burden of deeply nested properties. Timing Operator Error and Conversion / Type-check Error become more visible in higher tiers, reflecting that complex temporal operators and typing constraints are harder to satisfy in a single pass. Compared with general-purpose LLM baselines, the proposed approach exhibits consistently lower SPR error rates in lower tiers and remains competitive in higher tiers, indicating that SFT improves syntax robustness without relying on the regenerate loop.

\subsubsection{SER error attribution}
We report the distribution over four mutually exclusive relationships: equivalence; constraint widening; constraint tightening; and no relationship. Results are normalized over syntax-passed cases, so the four categories approximately sum to one for each model and tier. As difficulty increases from D1 to D4, the share of equivalence decreases for all models, while the share of no-relationship outcomes increases. This trend indicates that failures increasingly stem from semantic drift rather than small operator-level mistakes.

On D4, no relationship becomes the largest failure mode for the general-purpose LLMs, whereas the proposed approach maintains a substantially lower ``No Relationship'' rate. Specifically, the proposed approach is around 20\%, compared with roughly 29--32\% for the general-purpose LLMs, indicating better intent preservation under deep temporal nesting. When outputs are non-equivalent but still related, errors often manifest as directional implication: the generated property is either weaker (constraint widening) or stronger (constraint tightening) than the reference. These directional outcomes provide useful signals for targeted data refinement in the synthesis pipeline.

\section{Conclusion}
\label {sec:con}

This work presents a framework for automated SVA generation with LLMs, combining SFT-based assertion synthesis, AST-grounded supervision construction, and syntax-aware regeneration. On our depth-stratified benchmark, the proposed approach improves SER over the best general LLM by \textbf{+24.5} pp on D2, \textbf{+26.0} pp on D3, and \textbf{+17.5} pp on D4, yielding an average gain of \textbf{+22.7} pp over D2--D4 while maintaining comparable SPR. We further contribute an automated AST-grounded data construction pipeline and a benchmark suite with formal property equivalence checking. The benchmark suite will be released publicly upon publication.


\bibliographystyle{IEEEtran}
\bibliography{reference}

\end{document}